\documentclass[twocolappendix]{emulateapj}

\usepackage{amsmath,amssymb}

\usepackage{xspace}
\usepackage{graphics,graphicx}
\usepackage{natbib}
\usepackage{multirow}
\usepackage[percent]{overpic}
\usepackage{color}

\newcommand{\beq}{\begin{equation}}
\newcommand{\beqa}{\begin{eqnarray}}
\newcommand{\eeq}{\end{equation}}
\newcommand{\eeqa}{\end{eqnarray}}
\newcommand{\bfx}{\mbox{\boldmath{$x$}}}
\newcommand{\bfy}{\mbox{\boldmath{$y$}}}
\newcommand{\bfxi}{\mbox{\boldmath{$\xi$}}}
\newcommand{\bftheta}{\mbox{\boldmath{$\theta$}}}

\shorttitle{Arrival time differences between GWs and EM signals due to gravitational lensing}
\shortauthors{Takahashi}

\begin{document}

\title{Arrival time differences between gravitational waves and electromagnetic
  signals due to gravitational lensing}
\author{Ryuichi Takahashi}
\affil{Faculty of Science and Technology, Hirosaki University,
  3 Bunkyo-cho, Hirosaki, Aomori 036-8561, Japan}

\begin{abstract}
In this study, we demonstrate that general relativity predicts arrival time
differences between gravitational wave (GW) and electromagnetic (EM) signals
 caused by the wave effects in gravitational lensing.
The GW signals can arrive $earlier$ than the EM signals in some cases if
the GW/EM signals have passed through a lens, even if both signals were
emitted simultaneously by a source.
GW wavelengths are much larger than EM wavelengths; therefore, the propagation
of the GWs does not follow the laws of geometrical optics, including the Shapiro
time delay, if the lens mass is less than approximately $10^5 {\rm M}_\odot
(f/{\rm Hz})^{-1}$, where $f$ is the GW frequency.
The arrival time difference can reach $\sim 0.1 \, {\rm s} \, (f/{\rm Hz})^{-1}$
{ if the signals have passed by a lens of mass $\sim 8000{\rm M}_\odot (f/{\rm Hz})^{-1}$
  with the impact parameter smaller than the Einstein radius};
therefore, it is more prominent for lower GW frequencies.
{ For example, when a distant super massive black hole binary (SMBHB) in a galactic
  center is lensed by an intervening galaxy, the time lag becomes of the order of $10$ days.
  Future pulsar timing arrays including SKA (the Square Kilometre Array) and X-ray
  detectors may detect several time lags by measuring the orbital phase differences
  between the GW/EM signals in the SMBHBs.}
Gravitational lensing imprints a characteristic modulation on a chirp waveform;
 therefore, we can deduce whether a measured arrival time lag arises from intrinsic
 source properties or gravitational lensing.
Determination of arrival time differences would be extremely useful in multimessenger
 observations and tests of general relativity.
\end{abstract}


\section{Introduction}

The Laser Interferometer Gravitational-Wave Observatory (LIGO) recently reported the
first direct detection of gravitational waves (GWs) \citep{a16}.
The GW signal, named GW150914, comes from a binary black-hole (BH) merger with two
masses of $\sim 30 {\rm M}_\odot$ at a redshift of $\sim 0.09$.
Electromagnetic follow-up observations to GW150914 have been conducted over a wide
 wavelength range---from radio to $\gamma$-rays \citep{a16c,a16d}.
The electromagnetic (EM) signals provide crucial and complementary information concerning
 their source's properties.
Surprisingly, the Fermi Gamma-ray Burst Monitor (GBM) detected a hard X-ray
transit at $\sim 0.4$ s after the GW150914 trigger time, and this Fermi transit
may be associated with the GW transit\footnote{ However, the INTErnational Gamma-Ray
  Astrophysics Laboratory (INTEGRAL) did not detect a gamma-ray or a hard x-ray signal
  associated with it \citep{s16}.} \citep{c16,b16}.


There are several promising sources emitting both the GW and EM signals, such as supernovae,
 binary neutron star (NS) mergers, and NS--BH mergers.
The NS--NS/NS--BH mergers are usually considered to be associated with short gamma-ray
 bursts (GRBs).
BH--BH binaries also may emit EM signals if they are embedded in accretion disks.
Simultaneous observations of GWs, EM waves (EMWs), neutrinos, and cosmic rays are called
multimessenger astronomy, which is currently a very active field of research
\citep[see, e.g., a review by][]{a13,r15}.

Arrival time differences between GW and EM signals have been proposed as a test of 
 general relativity, as discussed below.
Orbital phase differences between the two signals in a white dwarf binary could be used to set
a constraint on the propagation speed of GWs and the graviton mass \citep{lh00,chl03,cs04}.
{ Similarly, orbital phase differences in a super massive BH binary (SMBHB) measured by
  pulsar timing arrays and optical telescopes can be used in the same manner \citep{k08}.}
Arrival time lags between the GW/EM signals from short GRBs or supernovae can also be used for
 the same purpose \citep{nn14,n16,l16}. 
In the latter case, we need to estimate the intrinsic emission time lags, which are usually
 unknown, on the basis of theoretical models.  
\cite{w16} and \cite{kd16} recently discussed a constraint on Einstein's equivalence principle
using the arrival time lag at different frequencies induced by the Shapiro time delay
resulting from intervening structures (but, they did not take into account the wave effects in
 gravitational lensing). 
\cite{a16b} and \cite{y16} discussed constraints on the GW propagation mechanism in detail
 referring to GW150914. 

In terms of multimessenger observations,
 one can ask a simple question. If a source simultaneously emits GWs and EMWs 
 and they propagate in a vacuum, will we detect them at the same time?  
We will demonstrate that the answer to this question is no, even in general relativity.
Gravitational lensing by an intervening object can generate a measurable arrival time
 difference.
 
The gravitational lensing of light is usually studied using the geometrical optics
approximation, because the wavelength of light is much smaller than typical lens sizes.
Maxwell's EM wave equation reduces to the null geodesic equation under this approximation
 \citep[][Chapter 3]{sef92}.
Then, when the light passes by the lens, it experiences the Shapiro time delay.
However, because the GW wavelength $\lambda$ is much larger,
 this approximation does not hold in some cases.
Roughly speaking, for lensing by a compact object with mass $M$, the geometrical optics
 approximation breaks down when $\lambda \gtrsim GM/c^2$ is satisfied, where $G$ is the
 gravitational constant and $c$ is the speed of light \citep[e.g.,][]{sef92,nd99}.
This can be rewritten as a condition for the lens mass
 $M \lesssim 10^5 {\rm M}_\odot (f/{\rm Hz})^{-1}$, where $f$ is the GW frequency.
This relation holds even for a non-Schwarzschild lens with a mass $M$, such as a galaxy,
 a star cluster, or a dark halo.
Similarly, for GW propagation in an inhomogeneous mass distribution, the
 geometrical optics approximation breaks down when $\lambda \gtrsim (k^2 D_{\rm S})^{-1}$,
 where $k$ is the wavenumber of the underlying density
 fluctuations and $D_{\rm S}$ is the distance to the source \citep[e.g.,][]{m04,t06}.
For such low GW frequencies, it is necessary to use wave optics instead of geometrical
 optics, and the GWs can pass by the lens without experiencing the Shapiro time delay
 or magnification.
Therefore, a wavelength difference can generate an arrival time difference. 


\section{Arrival Time Difference}

In this section, we will derive the arrival time difference based on wave optics in
 Section 2.1 and calculate it for simple lens models in Sections 2.2 and 2.3.

\subsection{Derivation of the Arrival Time Difference}

First, we briefly discuss previous studies on wave optics for the gravitational lensing
 of GWs \citep[see also a review by][]{sef92,nd99}. 
\cite{p74} was the first to study GW propagation in the weak gravitational field of a lens
object.
He derived a basic wave equation in curved spacetime and solved it exactly for a point mass
lens. 
\cite{o74} also derived a lensed waveform using the Kirchhoff diffraction integral, which
can be used for general lens mass distributions \citep[see also,][]{bm75,bh81}.
\cite{n98} discussed the diffraction effect on lensing magnifications for inspiraling binaries
induced by intervening compact objects.
Wave effects in the gravitational lensing of GWs have been discussed linked to several topics
\citep[e.g.,][]{b99,tn03,t04,tsm05,s05,s06,y07,s10,y13,c14}.

\begin{figure}[t]
\epsscale{1.1}
\plotone{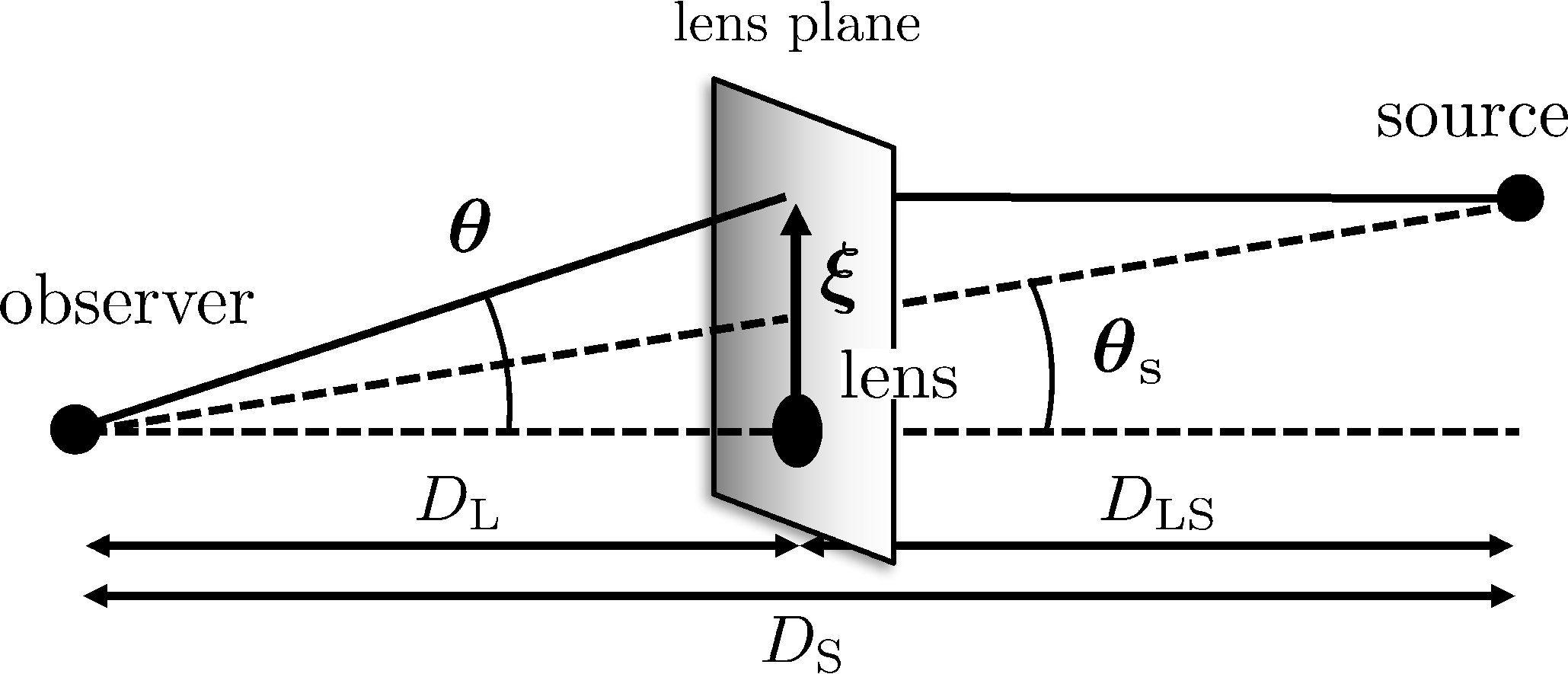}
\caption{The lensing configuration of an observer, a lens, and a source.
The angular diameter distances to the lens and the source are $D_{\rm L}$ and
$D_{\rm S}$, respectively. The distance between them is $D_{\rm LS}$.
The waves from the source are scattered on the lens plane at a point $\bfxi$.
The incoming wave direction is $\bftheta (=\bfxi/D_{\rm L})$, and
 the angular source position is $\bftheta_{\rm s}$.}
\label{fig1}
\end{figure}

Figure \ref{fig1} shows a lensing configuration with an observer, a lens, and a source
 in a nearly straight line.
The angular diameter distances to the lens and the source are $D_{\rm L}$ and $D_{\rm S}$,
respectively. The distance between them is $D_{\rm LS}$.
The lens mass distribution is projected on the lens plane (the thin-lens approximation),
 and the lens center is located at the origin ($\bftheta=0$) on the plane.
The waves from the source are scattered on the plane at a point $\bfxi$.
Then, the incoming direction of the wave is $\bftheta(=\bfxi/D_{\rm L})$,
while the angular source position is $\bftheta_{\rm s}$.
In geometrical optics, the light ray paths are uniquely determined by Fermat's principle
 (or the lens equation), while in wave optics, the GWs are described by the superposition
 (or integration) of many paths.
Throughout this study, we will discuss the lensing of light using geometrical optics
 and the lensing of GWs using wave optics.
In this section, we assume that the GWs are monochromatic with a frequency of $f$.
In Section 4, we will discuss a non-monochromatic case (a chirp signal).

The gravitationally lensed waveform of GWs is usually studied in the frequency
 domain (Fourier space) because the waveform has a simpler form than in the time domain.
 The waveform is a complex (real) function 
 in the frequency (time) domain. According to the previous studies,
the lensed waveform $\tilde{h}^{\rm L}_{+,\times}(f)$ in the frequency domain
is the product of a complex function $F(f,\bftheta_{\rm s})$ and the unlensed waveform
$\tilde{h}_{+,\times}(f)$, 
\beq
\tilde{h}^{\rm L}_{+,\times}(f) = F(f,\bftheta_{\rm s}) \tilde{h}_{+,\times}(f).
\label{lwf}
\eeq
The function $F$ is called the amplification factor or the transmission factor and
 is given by the diffraction integral \citep{sef92}:
\beq
F(f,\bftheta_{\rm s}) = \frac{D_{\rm L} D_{\rm S}}{c D_{\rm LS}} \frac{(1+z_{\rm L})f}{i}
\int d^2 \theta \exp \left[ 2 \pi i f t_{\rm d}(\bftheta,\bftheta_{\rm s}) \right],
\label{ampf}
\eeq
with the time delay,
\beq
 t_{\rm d}(\bftheta,\bftheta_{\rm s})= \frac{\left( 1+z_{\rm L} \right)}{c}
 \left[ \frac{D_{\rm L} D_{\rm S}}{2 D_{\rm LS}} \left| \bftheta-\bftheta_{\rm s} \right|^2
   - \frac{\hat{\psi}(\bftheta)}{c^2} \right],
\label{time_delay}
\eeq
relative to the unlensed case.
The first term is the geometrical time delay and the second term is the Shapiro
 time delay (or the potential time delay).
The overall factor $(1+z_{\rm L})$ is the cosmological time dilation, where $z_{\rm L}$ is
 the lens redshift.
The two-dimensional lens potential $\hat{\psi}(\bftheta)$ is determined by
 the surface mass density of the lens $\Sigma(\bftheta)$ via the Poisson equation,
 $\nabla^2_\theta \hat{\psi}(\bftheta) = 8 \pi G D^2_{\rm L} \Sigma(\bftheta)$.
The amplification factor $F$ is normalized to unity for the unlensed case $\hat{\psi}=0$.

We can rewrite the equations in their dimensionless forms by introducing a characteristic
 angular scale $\theta_*$, which will be chosen to be the angular Einstein radius.
We can define the { scaled} angle $\bfx=\bftheta/\theta_*$, source position
$\bfy=\bftheta_{\rm s}/\theta_*$, { the dimensionless} frequency
$w=[D_{\rm S}D_{\rm L}/(c D_{\rm LS})] \theta_*^2 (1+z_{\rm L}) 2 \pi f$, and time delay
 $T(\bfx,\bfy)=[c D_{\rm LS}/(D_{\rm S}D_{\rm L})] \theta_*^{-2} (1+z_{\rm L})^{-1}
 t_{\rm d}(\bftheta,\bftheta_{\rm s})$.
Then, the amplification factor has a simple form:
\beq
F(w,\bfy) = \frac{w}{2 \pi i} \int d^2x \exp \left[ iwT(\bfx,\bfy) \right].
\label{amp_fac2}
\eeq
In the low-frequency limit ($w \rightarrow 0$), the factor $F$ approaches unity.
In the high-frequency limit ($w \rightarrow \infty$), the stationary points of
 $T(\bfx,\bfy)$ contribute to the integral in Eq.(\ref{amp_fac2}).
Then, multiple images can form and these angular positions can be determined by solving
 the lens equation, $\nabla_x T(\bfx,\bfy)=0$.
Given the $j$th image position $\bfx_j$, its time delay and magnification are
 $T_j=T(\bfx_j,\bfy)$ and $\mu_j = [ \det (\partial \bfy/\partial \bfx_j) ]^{-1}$,
 respectively.
Hereafter, $T_{{\rm EM},j}$ indicates the time delay of the $j$th EM image.

When the wavelength is larger than the path difference between the multiple images,
 the geometrical optics approximation breaks down \citep[e.g.,][]{sef92}.
This condition can be rewritten as $w(T_i - T_j) \lesssim 1$ for the $i$th and $j$th images.
Throughout this paper, we take $w(T_i-T_j)=1$ to be the border between the geometrical-optics
 regime and the wave-optics regime.
We assume that the GW wavelength is large enough to be in the wave-optics regime.
Then, there is always only one GW image.
We define the time delay of the GW signal from the phase of the amplification factor:
\beq
T_{\rm GW}(w,\bfy) \equiv - \frac{i}{w} \ln \left( \frac{F(w,\bfy)}{\left| F(w,\bfy)
  \right|} \right).
\label{td_gw}
\eeq
Note that the GW time delay depends on the frequency.

We can define the arrival time difference of the $j$th EM image signal relative to
 the GW signal as
\beq
\Delta T_{{\rm EM},j-{\rm GW}}(w,\bfy) \equiv T_{{\rm EM},j}(\bfy) - T_{\rm GW}(w,\bfy).
\label{td_diff}
\eeq
The time difference $\Delta T_{{\rm EM},j-{\rm GW}}$ definitely approaches zero in the
 high-frequency limit ($w \rightarrow \infty$), when there is a single EM image. 
However, it does not happen when there are multiple EM images.
This is because, as the GW frequency increases into the geometrical-optics regime,
 multiple GW images appear.
Then, the $T_{\rm GW}$ contains the time delays of several images, not just the $j$th image.

In the following subsections, we will calculate the arrival time differences in
 Eq.(\ref{td_diff}) for specific lens models: a point mass (Section 2.1)
 and a singular isothermal sphere (Section 2.2).
 { We will basically discuss in dimensionless form for convenience 
   but comment for specific lens systems (sources, lenses, and
   detectors) in dimensional form at the end of Section 2.2 and Section 3.}

\subsection{A Point Mass Lens}

\begin{figure*}
\hspace*{-1.5cm}
\begin{minipage}{\columnwidth}
\includegraphics[width=1.4\columnwidth]{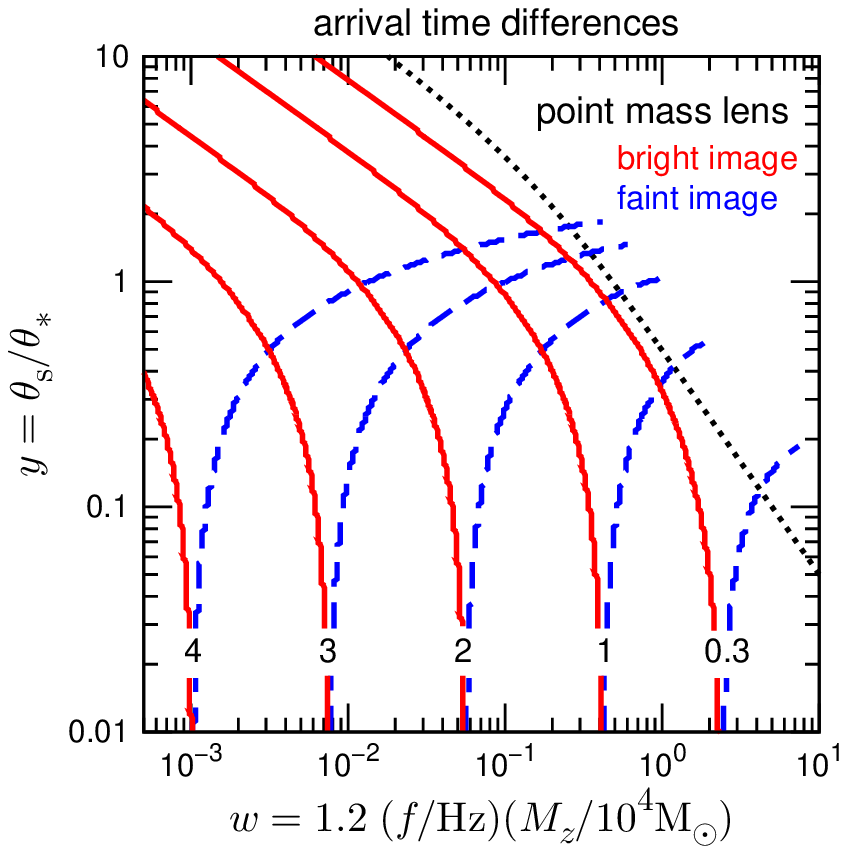}   
\end{minipage}
\begin{minipage}{\columnwidth}
\includegraphics[width=1.4\columnwidth]{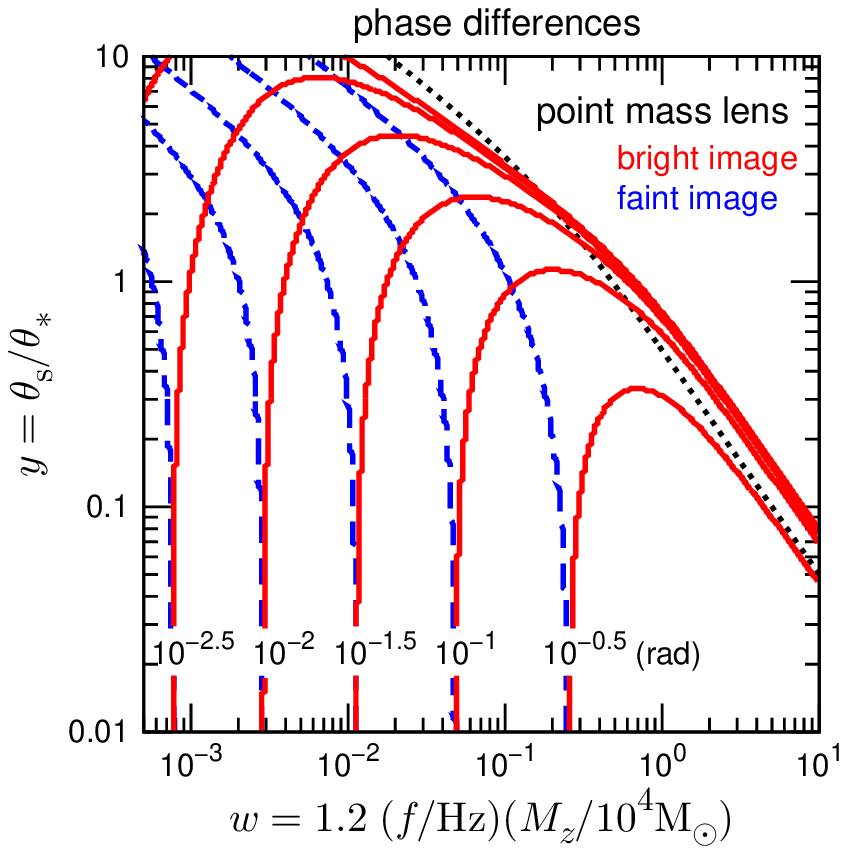}
\end{minipage}
\caption{{\bf Left Panel:} Contour plot of the arrival time differences of the bright
 (solid red,$+$) and the faint (dashed blue,$-$) EM image signals relative to the GW signal.
  The horizontal axis is the dimensionless GW frequency $w = 8 \pi (GM_z/c^3) f$
  $=1.2 (f/{\rm Hz}) (M_z/10^4{\rm M}_\odot)$, where $f$ is the GW frequency and
  $M_z=(1+z_{\rm L}) M$ is the redshifted lens mass.
  The vertical axis is the angular source position
  normalized by the Einstein angle, $y = \theta_{\rm s} / \theta_*$.
  The arrival time differences are $\Delta t_{{\rm d},{\rm EM},\pm-{\rm GW}} =
  4 (GM_z/c^3) \Delta T_{{\rm EM},\pm-{\rm GW}}$ $= 0.20 \, {\rm sec} \,
  \Delta T_{{\rm EM},\pm-{\rm GW}} (M_z/10^4{\rm M}_\odot)$, where $\Delta T_{{\rm EM},\pm-{\rm GW}}
  = 0.3,1,2,3,$ and $4$ are denoted as the solid red ($+$) and dashed blue ($-$) curves.
The dotted black curve denotes the boundary between the geometrical-optics regime
 (the top-right side) and the wave-optics regime (the bottom-left side).
Please look at the bottom-left side.
The positive time differences indicate that the GW signal arrives earlier than the EM signals.
 {\bf Right Panel:} Same as left panel, but for the phase differences 
 $2 \pi f \Delta t_{{\rm d},{\rm EM},\pm-{\rm GW}} (= w \Delta T_{{\rm EM},\pm-{\rm GW}})$
 with contour levels at $10^{-2.5},10^{-2},10^{-1.5},10^{-1},$ and $10^{-0.5}$ rad.
}
\label{fig_pm}
\end{figure*}

First, we consider lensing by a compact object (for instance, a star, planet,
 or black hole) with mass $M$.
The surface density profile and the potential are $\Sigma(\theta)$ $=M
 \delta^{2}(D_{\rm L}
 \bftheta)$ and $\hat{\psi}(\theta)$ $=4 G M \ln \theta$, respectively.
We chose the angular Einstein radius to be the characteristic scale
\beq
 \theta_* = \sqrt{\frac{4 G M D_{\rm LS}}{c^2 D_{\rm L} D_{\rm S}}}.
\nonumber
\eeq
Then, the source position, { the dimensionless} GW frequency, and time delay are
\beq
y=\frac{\theta_{\rm s}}{\theta_*}, ~w=8 \pi \frac{GM_z}{c^3} f, ~{\rm and}~
  T(\bfx,\bfy)= \frac{c^3}{4GM_z} t_{\rm d}(\bftheta,\bftheta_{\rm s}),
\label{ywT}
\eeq
respectively.
Here, $M_z=(1+z_{\rm L})M$ is the redshifted lens mass.
The four parameters of the lens mass $M_z$, source position $\theta_{\rm s}$,
  GW frequency $f$, and time delay $t_{\rm d}$ reduce to the three dimensionless
  parameters of $y,w,$ and $T$.
The amplification factor is
 $F(w,y) = \exp [{\pi w}/{4} + ({iw}/{2}) \ln ({w}/{2})]$
 $\Gamma (1-{iw}/{2})$ ${}_1F_1 ( {iw}/{2},1 ; {iwy^2}/{2})$,
where ${}_1F_1$ is the the confluent hypergeometric function \citep{p74}.
For very low frequency ($w \ll 1$), we can obtain the GW time delay in Eq.(\ref{td_gw})
 as a series in $w$,
\beq
T_{\rm GW}(w,y)=\frac{1}{2} \left[ \ln \left( \frac{w}{2} \right) + \gamma \right]
+\mathcal{O}(w^2),
\label{appr_tgw}
\eeq
where $\gamma=0.5772\cdots$ is the Euler constant.
This GW time delay is independent of the source position $y$.


In the geometrical-optics limit, the two EM images always form a bright ($+$) image
 and a faint ($-$) image.
Their time delays and magnifications are $T_{{\rm EM},{\pm}}(y)$$=(y^2+2 \mp y \sqrt{y^2+4})/4
- \ln|(y \pm \sqrt{y^2+4})/2|$ and $\mu_{{\rm EM},{\pm}}$$=1/2 \pm (y^2+2)/(2y\sqrt{y^2+4})$,
 respectively.

Figure \ref{fig_pm} (left panel) shows the arrival time differences
 $\Delta T_{{\rm EM},\pm-{\rm GW}}(w,y)$ in Eq.(\ref{td_diff}) as a contour map.
The horizontal axis is the dimensionless GW frequency $w = 8 \pi (GM_z/c^3) f$
 $=1.2 (f/{\rm Hz}) (M_z/10^4{\rm M}_\odot)$, while the vertical axis is
 the source position $y=\theta_{\rm s}/\theta_*$.
The solid red and dashed blue curves show the results for the bright ($+$) and
 faint ($-$) EM images, respectively.
The results simply scale in proportion to the lens mass $M_z$ via
 $w \propto M_z f$ and $\Delta t_{\rm d} \propto M_z \Delta T$.
The dotted black curve denotes the boundary between the geometrical- and the wave-optics
 regimes at $w (T_{+}-T_{-})=1$.
The top-right (bottom-left) side of the dotted curve is the geometrical-(wave-)optics regime.
{ In plotting the figure, we used the exact formula in Eq.(\ref{td_gw}), not the
  approximate one in Eq.(\ref{appr_tgw}).}
As expected, the time differences are larger for lower frequencies.
As the dotted curve (the boundary) is approached, the solid-red contours approach 
 zero; however, the dashed-blue contours do not because the GW time delay contains the
 contributions of both images and the bright image is dominant.
For a very small source position $y \ll 1$, the time delays of the bright and
 faint images are asymptotically the same; therefore, the solid red and dashed
 blue curves converge. 

As seen in the figure, the time differences are always positive, which means that the GW
 signal arrives earlier than both of the EM signals.
This may seem surprising and counter-intuitive because the bright EM image corresponds
 to the minimum time delay via Fermat's principle.
To understand why the GWs arrive earlier, we can rewrite the amplification factor,
 Eq.(\ref{ampf}), with the time delay, Eq.(\ref{time_delay}), by replacing the variables
$\bftheta^\prime=\sqrt{(1+z_{\rm L}) f}\bftheta$ and
$\bftheta_{\rm s}^\prime= \sqrt{(1+z_{\rm L}) f}\bftheta_{\rm s}$,
\beqa
  && F(f,\bftheta_{\rm s})=\frac{D_{\rm L}D_{\rm S}}{i c D_{\rm LS}} \int d^2 \bftheta^\prime
  \exp \left[ \frac{2 \pi i}{c} \left\{ \frac{D_{\rm L}D_{\rm S}}{2 D_{\rm LS}}
  \left|\bftheta^\prime - \bftheta_{\rm s}^\prime \right|^2 \right. \right. \nonumber \\
  && \hspace{2.5cm}
  \left. \left.  - \frac{(1+z_{\rm L}) f}{c^2} \hat{\psi}
  \left( \frac{\bftheta^\prime}{\sqrt{(1+z_{\rm L})f}}
  \right) \right\} \right].
\eeqa
Therefore, in the low-frequency limit $(f \rightarrow 0)$, the potential term is
 negligibly small.
Then, a point near $\bftheta^\prime=\bftheta^\prime_{\rm s}$ contributes to the integral,
 which means that the GWs pass in a straight line from the source to the observer
 and therefore the geometrical time delay (the first term in Eq.(\ref{time_delay})) is
 also small, and the GW time delay is smaller than both of the EM time delays.

Even for a relatively high source position $y>1$, an arrival time
 difference arises for the bright image, as seen in Fig.\ref{fig_pm} (left panel).
This is because $\Delta T_{{\rm EM},\pm-{\rm GW}}$ is not sensitive to $y$ because
(i) the Shapiro time delay is proportional to the logarithm of $y$ and
(ii) the GW time delay is independent of $y$ { in the low frequency limit from
  Eq.(\ref{appr_tgw})}.

In GW measurements, the time difference can be measured from the phase difference.
Therefore, we plot the phase differences defined by $w \Delta T_{{\rm EM},\pm-{\rm GW}}$
in Fig.\ref{fig_pm} (right panel).
As seen in the figure, the phase differences are largest at $w \simeq 1$ with
 $y \lesssim 0.3$.
The maximum phase differences are $w \Delta T_{{\rm EM},\pm-{\rm GW}}^{\rm max}$
$\simeq 0.1, 0.5,$ and $0.7$ rad for $y$ $=1,0.1,$ and $0.01$, respectively.
In a matched filtering analysis, the phase of the waveform can be roughly measured
within the accuracy of the inverse signal-to-noise ratio $\approx
({\rm S}/{\rm N})^{-1}$ \citep{cf94}.\footnote{Here, we ignore the measurement
    accuracies of the EM arrival times which are assumed to be determined with
    infinite accuracy.
 { We will mention the real time lags for specific lens systems later.}}
For example, we can measure the phase differences for a lensed GW signal with
  ${\rm S}/{\rm N}=10 (100)$ if the parameters $w$ and $y$ are
  in the region where $w \Delta T_{{\rm EM},\pm-{\rm GW}} \gtrsim 10^{-1} (10^{-2})$ rad
  in the right panel.
The arrival time differences can be most accurately determined near $w \simeq 1$.

Finally, let us comment on the ``dimensional'' arrival time difference.
The dimensional time delay and frequency are proportional to the dimensionless
 quantities via $t_{\rm d} \propto M T$ and $f \propto w/M$.
Therefore, we obtain $t_d \propto w T$ for the fixed frequency $f$.
From the phase relation $2 \pi f \Delta t_{{\rm d},{\rm EM},\pm-{\rm GW}}=w \Delta
 T_{{\rm EM},\pm-{\rm GW}}$, we can estimate the dimensional time difference as
\beqa
&& \Delta t_{{\rm d},{\rm EM},\pm-{\rm GW}}
 = \frac{1}{2 \pi f} w \Delta T_{{\rm EM},\pm-{\rm GW}}, \nonumber   \\
&& \hspace{1.2cm} = 0.16 \, {\rm sec} \left( \frac{f}{{\rm Hz}} \right)^{-1} 
\left( \frac{w \Delta T_{{\rm EM},\pm-{\rm GW}}}{1} \right). 
\label{delta_td_wT}
\eeqa
Therefore, the dimensional time difference 
 is more prominent for lower GW frequencies and larger phase differences.
The phase difference is largest at $w \simeq 1$ and the corresponding lens
  mass is
\beqa
M_z &=& \frac{c^3}{8 \pi G} \frac{w}{f},  \nonumber \\
 &=& 8100 {\rm M}_\odot \left( \frac{f}{{\rm Hz}} \right)^{-1} \left( \frac{w}{1} \right),
\label{lens_mass}
\eeqa
from Eq.(\ref{ywT}).
Because the maximum phase difference is $w \Delta T^{\rm max}_{{\rm EM},\pm-{\rm GW}}
 \simeq 0.7$, the maximum time difference is
\beq
\Delta t_{{\rm d},{\rm EM},\pm-{\rm GW}}^{\rm max} \simeq 0.11 \, {\rm sec} \,
\left( \frac{f}{\rm Hz} \right)^{-1},
\label{delta_td_max}
\eeq
from Eq.(\ref{delta_td_wT}).
This is typically $\sim 1$ ms $(f/100{\rm Hz})^{-1}$ for ground-based
 detectors, $\sim 2$ min $(f/{\rm mHz})^{-1}$ for space-based interferometers,
 and $\sim 4$ months $(f/10^{-8}{\rm Hz})^{-1}$ for pulsar timing arrays.

{
For the ground-based detectors, extra-galactic compact binaries (NS--NS/NS--BH)
and nearby supernovae are promising sources which may be associated with EM signals.
Similarly, for the planned space interferometer eLISA (the Evolved Laser Interferometer
Space Antenna), galactic white dwarf binaries are the most secure sources.
The orbital phase differences between the GW/EM signals in these binaries can be used
 to measure the arrival time differences.
Here let us consider that the sources are lensed by galactic stars as a plausible case.
Then, the lensing-induced time lag is typically
 $\sim (GM/c^3) \sim 10^{-5} \, {\rm s} \, (M/{\rm M}_\odot)$ but it 
 is too small to be resolved by optical detectors.
The angular separation between the bright/faint images is
approximately $10^{-6}$ deg which is also very small.
The lensing probability that the GW/EM signals from the source pass near the star within
the Einstein radius is $\approx 10^{-6} (\lesssim 10^{-8})$ for the direction to the
Galactic bulge (perpendicular to the Galactic disk) \citep[e.g.][]{ws11}.
Therefore, detecting the time lags in such lens systems in the near future is unlikely.
}

\subsection{A Singular Isothermal Sphere Lens}

\begin{figure*}
\hspace*{-1.5cm}
\begin{minipage}{\columnwidth}
\includegraphics[width=1.4\columnwidth]{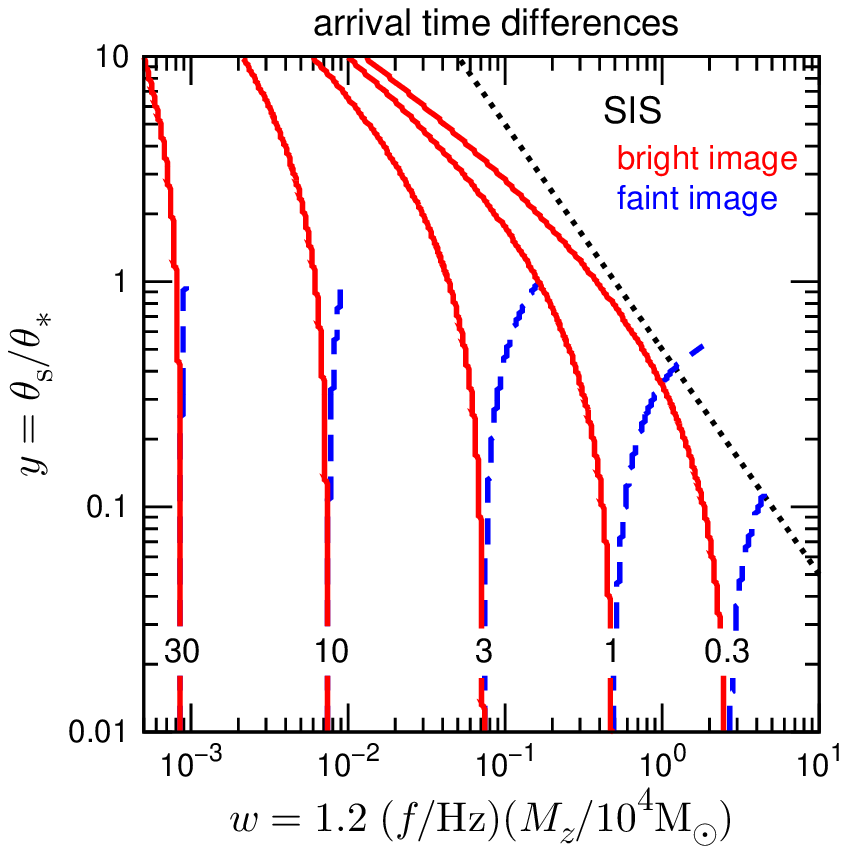}   
\end{minipage}
\begin{minipage}{\columnwidth}
\includegraphics[width=1.4\columnwidth]{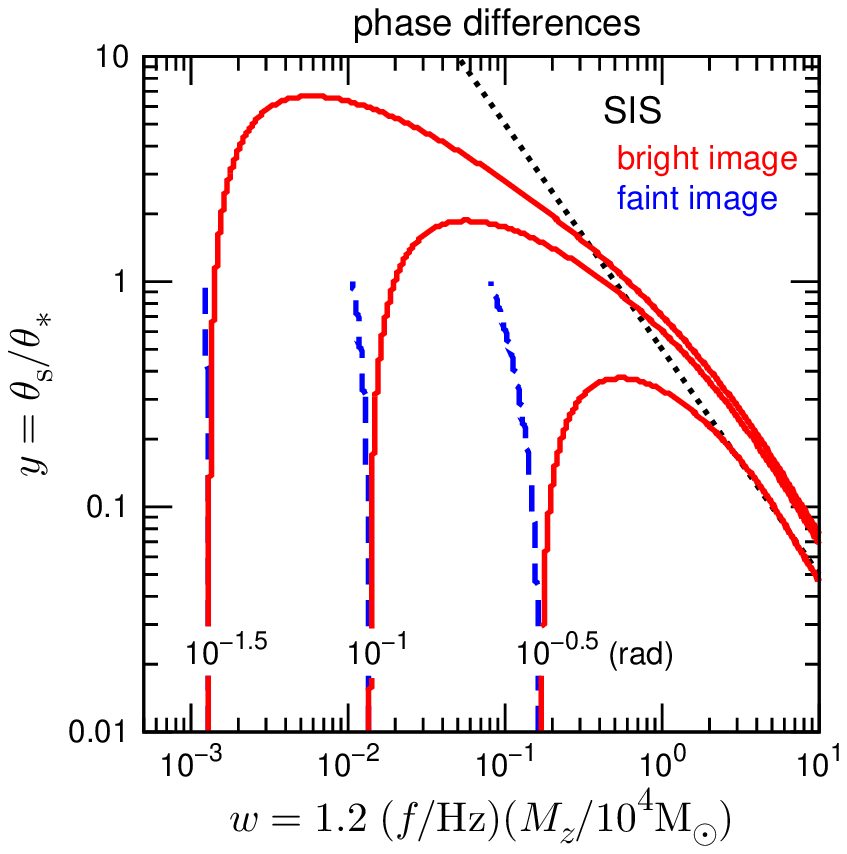}
\end{minipage}
\caption{ Same as Fig.\ref{fig_pm}, but for the SIS (singular isothermal sphere) lens.
  The faint image only appears for $y<1$.
  {\bf Left Panel:} The time differences are $\Delta t_{{\rm d},{\rm EM},\pm-{\rm GW}}$
  $= 0.20 \,
  {\rm sec} \, \Delta T_{{\rm EM},\pm-{\rm GW}} (M_z/10^4{\rm M}_\odot)$,
  where $\Delta T_{{\rm EM},\pm-{\rm GW}} = 0.3,1,3,10$ and $30$ are denoted as the curves.
  {\bf Right Panel:} The phase differences $2 \pi f \Delta t_{{\rm d},{\rm EM},\pm-{\rm GW}}$
  $(= w \Delta T_{{\rm EM},\pm-{\rm GW}})$ are shown with
  contour levels at $10^{-1.5},10^{-1}$ and $10^{-0.5}$ rad.
}
\label{fig_sis}
\end{figure*}

Next, we consider lensing by a singular isothermal sphere (SIS), which 
 approximately represents the density profile of a galaxy, dark halo, or star cluster.
The surface density is characterized by the velocity dispersion $v$ as
$\Sigma(\theta)=v^2/(2 G D_{\rm L} \theta)$. The potential is
$\hat{\psi}(\theta)=4 \pi v^2 D_{\rm L} \theta$.
We adopt the Einstein radius as the characteristic scale, $\theta_*=4 \pi (v/c)^2
D_{\rm LS}/D_{\rm S}$.
Then, the dimensionless GW frequency and time delays are the same as in the point mass
 lens with a mass of $M_{z}=4 \pi^2 (v^4/(c^2 G)) (1+z_{\rm L}) D_{\rm L}
 D_{\rm LS}/D_{\rm S}$, which is the enclosed mass within the Einstein radius.
Next, we can obtain the amplification factor $F(w,y)$ via numerical integration
\citep[e.g.,][]{tn03,m08}.
The GW time delay for $w \ll 1$ is
\beq
T_{\rm GW}(w,y) = -\frac{\sqrt{\pi}}{2} w^{-{1}/{2}} -\left( 1- \frac{\pi}{4} \right)
 +\mathcal{O}(w^{1/2}),
 \eeq
which is independent of $y$.

In geometrical optics, two images or one image forms for $y<1$ or $y \geq 1$,
 respectively.
The time delays and magnifications are $T_{\pm}=\mp y - 1/2$ and $\mu_{\pm}=\pm 1 +1/y$,
 respectively, for the bright ($+$) and faint ($-$) images.
The faint image appears only for $y<1$.

Figure \ref{fig_sis} shows the arrival time differences { and the phase differences}
for the SIS lens.
The dotted line is extrapolated to $y \geq 1$.
Similar to the results of Fig.\ref{fig_pm}, the GWs can arrive before both EM images. 
The time differences are larger than those of the point mass lens.




\section{ Lensing Event Rate for Future PTAs}

\begin{figure*}[t]
\vspace*{-3cm}
\epsscale{1.1}  
\plotone{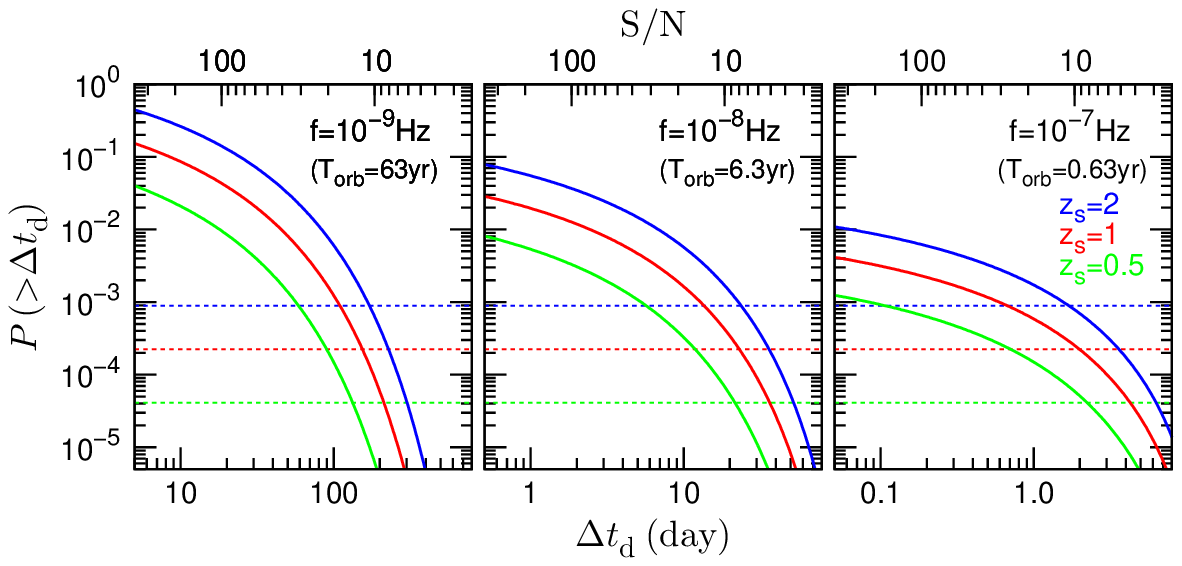}
\caption{
  { The lensing probability that the arrival time difference of the GW/EM signals
    from a distant BH binary is larger than $\Delta t_{\rm d}$ due to lensing by a
    line-of-sight galaxy.
    The GW frequencies are $f=10^{-9}$Hz (left panel), $f=10^{-8}$Hz (middle panel),
    and $f=10^{-7}$Hz (right panel).
    The orbital periods are shown in the parentheses ($T_{\rm orb}=2/f$). 
    The lower-x axes are the arrival time lags (in units of days) between the bright EM
    signals and the GW signals,
    while the upper-x axes are the signal-to-noise ratios necessary
    to resolve the time lags.
    The solid curves are the results at source redshifts $z_{\rm s}=2$ (blue),
    $z_{\rm s}=1$ (red), and $z_{\rm s}=0.5$ (green).
  The horizontal dashed lines are the lensing probabilities to form the double source images.}
}
\label{fig_lp2}
\end{figure*}

{
  In this section, we calculate the probability that a distant SMBHB
  at a galactic center is lensed by a line-of-sight galaxy for future PTAs.
As discussed in Section 2, we can measure the arrival time differences more
 easily for lower GW frequencies.
Therefore the PTAs are the most suitable to detect them. 
For the PTA frequency band ($f \approx 10^{-8}$ Hz), the typical lens mass
(for which the time difference arises most efficiently)
is the galactic scale 
 $M_z \simeq 8 \times 10^{11} {\rm M}_\odot (f/10^{-8} {\rm Hz})^{-1} (w/1)$
from Eq.(\ref{lens_mass}), and the maximum time difference is about a few months.
  
The PTAs aim to detect ultra low-frequency GWs in a range of $10^{-9}-10^{-6}$Hz.
Currently the European PTA, the Parks PTA, and NANOGrav (the North American Nanohertz
Observatory for Gravitational Waves) are in operation
\citep[see, e.g., a review by][]{h10,b15}.
Furthermore, the Square Kilometre Array (SKA) will start operating in 2020 \citep{j15}.
The SMBHBs in centers of merging galaxies are prime targets for the PTAs.
In such low frequencies, 
 the SMBHBs are long before the coalescence and the orbital motions are stationary.
GW signals from the binaries form a stochastic background, but some massive
($M>10^8 {\rm M}_\odot$) and near ($z_{\rm s}<2$) sources can be individually resolved
\citep{svv09}.
\cite{s12} estimated that the future PTAs in the SKA era will detect about $100-500$
resolved SMBHBs at $z \simeq 0.2-2$ in both GW and X-ray signals assuming that all
the binaries have circumbinary discs.
Here, we assume that the EM signals trace the orbital motions of the binaries
 and we can measure the arrival time differences from the
 orbital phase differences between the GW/EM signals (see also \cite{k08}, Section 4,
 for a similar argument).
We calculate the probability that the GW/EM waves pass near the lens galaxy.
We assume that the lens galaxy is simply modeled as the SIS.
The lensing probability is roughly estimated as
\beq
P(>\Delta t_{\rm d}) \sim \sigma(\Delta t_{\rm d}) n D_{\rm S},
\label{lp_estimate}
\eeq
where $\sigma(\Delta t_{\rm d}) \sim \pi r^2_{\rm E} y^2$ is the cross section that
the time lag is larger than $\Delta t_{\rm d}$, and $n$ is the number density
of the lens.
We adopt the velocity function of SDSS galaxies \citep{b10} for the number density $n$.
We present the detailed derivation of the probability distribution of $\Delta t_{\rm d}$
 in Section 3.1.

\subsection{Lensing Probability for Future PTAs}

In this subsection, we present our detailed calculation of the lensing probability that
 a distant SMBHB is lensed by a foreground galaxy.
We suppose that the future PTAs will detect the GW signals.
Here, we basically follow the theoretical model in \cite{o12} to evaluate the lensing
 probability.
They calculated the probability that a distant quasar is lensed by
 a foreground galaxy to form multiple images.
Their model agrees well with observational results in the SDSS Quasar Lens Search (SQLS).
We assume that the galaxy density profile is described by the SIS model and
the source at redshift $z_{\rm S}$ emits the GWs with frequency $f$.
Then, the lensing probability that the time-lag between the EM/GW signals is larger than
$\Delta t_{\rm d}$ is written as, 
\beq
P(> \! \Delta t_{\rm d};f,z_{\rm S}) = \int \! cdt \int \!
 d\sigma(\Delta t_{\rm d},f,z_{\rm S}) \int \! dn,
\eeq
where $cdt=c(dt/dz_{\rm L})dz_{\rm L}=- c dz_{\rm L}/(H(z_{\rm L}) (1+z_{\rm L}))$ where
$H(z_{\rm L})$ is the Hubble expansion rate at the lens redshift $z_{\rm L}$.
We adopt the cosmological parameters of the Hubble parameter
$H_0=70 \, {\rm km s^{-1}} \, {\rm Mpc}^{-1}$, the matter density $\Omega_{\rm m}=0.27$,
and the cosmological constant $\Omega_{\Lambda}=0.73$. 
The cross section is,
\beqa
d\sigma(\Delta t_{\rm d},f,z_{\rm S}) &=& 2 \pi r_{\rm E}^2 dy y
\nonumber \\
&& \times \Theta ( \Delta t_{\rm d} - \Delta t_{{\rm d,EM},\pm-{\rm GW}}(w,y) ),
\eeqa
where the Einstein radius is $r_{\rm E}=4 \pi (v/c)^2 D_{\rm L} D_{\rm LS}/D_{\rm S}$
and the step function is $\Theta(x)=1(0)$ for $x \geq 0 (x<0)$.
The arrival time difference between the GW signal and the bright/faint EM image ($\pm$)
signals $\Delta t_{{\rm d,EM},\pm-{\rm GW}}(w,y)$ is numerically evaluated as in Section 2.3. 
The number density of the lens is $dn=(dn/dv)dv$ where $dn/dv$ is the velocity-dispersion
 function of galaxies.
Suppose that the galaxy velocity function does not evolve in time, then it has a
 simpler form, $dn(v,z_{\rm L})/dv=
 (1+z_{\rm L})^3 dn_0/dv(v)$ where $dn_0/dv$ is the present velocity function. 
We adopt the velocity function in SDSS data \citep{b10}.
They provided a distribution function of velocity dispersion for all types
 (early and late types) galaxies using $\sim 2.5 \times 10^5$ samples:
\beq
\frac{dn_0}{dv}(v) = \frac{\phi_*}{v} \frac{\beta}{\Gamma(\alpha/\beta)}
\left( \frac{v}{v_*} \right)^\alpha
\exp \left[ - \left( \frac{v}{v_*} \right)^\beta \right] 
\eeq
with $\alpha=0.07$, $\beta=2.22$, $v_*=141 {\rm km/s}$, and $\phi_*=2.47 \times 10^{-1}
 {\rm Mpc^{-3}}$.
Here, $\Gamma(x)$ is the Gamma function.

Let us estimate the typical velocity dispersion which cause the maximum time-lag for
the PTA frequency band ($f=10^{-9}-10^{-7}$Hz).
Using Eq.(\ref{lens_mass}) with the mass enclosed within the Einstein radius
 ($M_z = 4 \pi^2 (v^4/(c^2 G)) (1+z_{\rm L}) D_{\rm L} D_{\rm LS}/D_{\rm S}$),
the typical velocity dispersion is
\beqa
v = 298 ~{\rm km} \, {\rm s}^{-1} \, \left( 1+z_{\rm L} \right)^{-1/4}
\left( \frac{D_{\rm L}D_{\rm LS}/D_{\rm S}}{\rm Gpc} \right)^{-1/4} \nonumber \\ 
\times \left( \frac{f}{10^{-8} {\rm Hz}} \right)^{-1/4}
\left( \frac{w}{1} \right)^{1/4}. 
\eeqa
The above velocity dispersion is comparable to $v_*$ in the velocity function.
Therefore, the galaxies are the suitable lenses for the PTAs.

Finally, we have the lensing probability as
\beqa
&& P(> \! \Delta t_{\rm d};f,z_{\rm S}) = 32 \pi^3 \int_0^{z_{\rm s}} \!\!
\frac{c dz_{\rm L}}{H(z_{\rm L})} (1+z_{\rm L})^2 \left( \frac{D_{\rm L} D_{\rm LS}}{D_{\rm S}}
\right)^2  \nonumber \\
 && ~~~~~~  \times  \int_0^\infty \!\! dy y \int_0^\infty \!\! dv \frac{dn_0}{dv}(v)  
\left( \frac{v}{c} \right)^4 \nonumber \\
 && ~~~~~~  \times \Theta( \Delta t_{\rm d} -
 \Delta t_{{\rm d,EM},\pm-{\rm GW}}(w,y)).
\eeqa
When calculating the probability to form the double source images, we simply replace
 the step function to $\Theta(1-y)$ in the above equation.
 
\subsection{Results}

Fig.\ref{fig_lp2} shows the probability that the arrival time difference is larger
 than $\Delta t_{\rm d}$.
The lower-x axes are the time lags $\Delta t_{\rm d}$ between the bright EM and the GW
signals, while the upper-x axes are the S/N necessary to measure the time lags
(which is simply ${\rm  S/N}=(2 \pi f \Delta t_{\rm d})^{-1}$).
The GW frequencies are $f=10^{-9}$Hz to $f=10^{-7}$Hz from left to right.
The corresponding orbital periods are $T=2/f=63$yr, $6.3$yr and $0.63$yr, respectively.
The source redshifts are $z_{\rm s}=2,1,$ and $0.5$ from top to bottom.
The horizontal dashed lines are the probabilities that the double source images form 
(note again that they form when $y<1$).
As shown in the figure, the solid curves are much higher than the dashed lines for
 smaller $\Delta t_{\rm d}$.
This is because for high S/N the smaller phase difference can be resolved even for $y>1$
 (see right panel of Fig.\ref{fig_sis}).
Then, the cross section is proportional to $y^2$ and thus the lensing probability
becomes larger.
For instance, the maximum $y$ to resolve the signal is about $2 (20)$ for
${\rm S/N}=10 (100)$.
In such a high $y (>1)$, only the bright EM image appears.
From the figure, for $f=10^{-8}$Hz at $z_{\rm s}=1$, the lensing
 probability of $\Delta t_{\rm d}>2(10)$ days is $1 \times 10^{-2}$ $(2 \times10^{-3})$
 with the necessary S/N of $100(20)$.
Therefore, if some hundred sources will be resolved as suggested in \cite{s12}, some
 lensing events will be detected. 

At present, over $140$ lensed quasar-galaxy systems are known\footnote{These systems
  are tabulated in following web sites: CASTLES Survey
  (https://www.cfa.harvard.edu/castles/)
  and the Sloan Digital Sky Survey Quasar Lens Search
  (http://www-utap.phys.s.u-tokyo.ac.jp/\~{}sdss/sqls/).}.
In the near future the Large Synoptic Survey Telescope (LSST)
 \footnote{LSST web page:https://www.lsst.org/} will find $\sim 8000$
 lensed quasars \citep{om10}.
If we find SMBHBs at the centers of such lensed quasars, we will possibly 
 detect the time lags in such systems.
}


\section{The Lensed Chirp Waveform}

\begin{figure*}
\epsscale{1.2}
\plotone{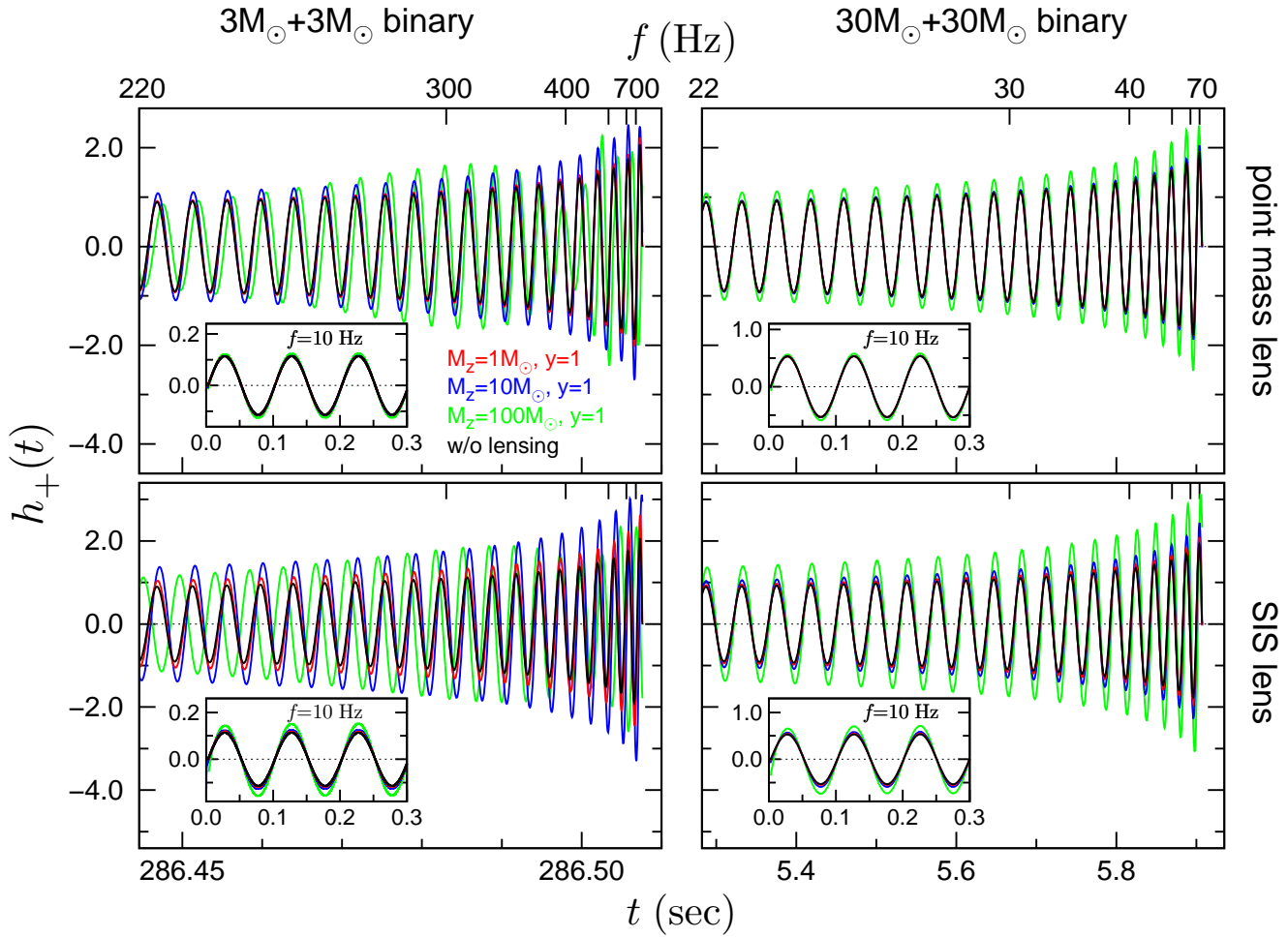}
\caption{The inspiral waveforms for the $3{\rm M}_\odot+3{\rm M}_\odot$ BH binary
  (left panels) and the $30{\rm M}_\odot+30{\rm M}_\odot$ BH binary (right panels)
  lensed by a point mass lens (upper panels) and by an SIS lens (lower panels).
  The lower x-axes show the time elapsed since the initial frequency $f=10$ Hz, and the
  upper x-axes show the corresponding frequency.
  The y-axes are the amplitudes with arbitrary units.
  The red, blue, and green curves are the lensed waveforms with the lens masses
  $M_z=1{\rm M}_\odot, 10{\rm M}_\odot$ and $100 {\rm M}_\odot$ with a fixed source
  position $y=1$. The black curves are the unlensed waveforms.
  The lower-left subpanels show the initial waveforms during the first $0.3$ s (when the
  frequency is $f=10$ Hz).}
\label{lensed_waveforms}
\vspace*{0.5cm}
\end{figure*}

If we detect GW and EM signals from the same source at different times, how can we  
 know whether the arrival time difference is due to an 
 intrinsic emission time lag or gravitational lensing?
We can, in principle, distinguish the two by measuring the frequency dependence of the
 time delay.
The GW time delay has a characteristic frequency dependence.
Furthermore, the lensing magnification also depends on the GW frequency.
In the wave-optics regime, the magnification approaches unity in the low-frequency
 limit due to the diffraction effect \citep[e.g.,][]{n98,tn03}.
The GW frequency from an inspiraling binary sweeps from low to high (the so called
 chirp signal), and therefore we can measure the frequency dependence in the GW waveform.
In this section, we will demonstrate the lensing effects on the chirp waveforms of
 inspiral BH binaries.

Consider two BH binary sources with equal masses $3{\rm M}_\odot+3{\rm M}_\odot$ and
 $30{\rm M}_\odot+30{\rm M}_\odot$.
We can adopt the Taylor T4 formula for the inspiral waveforms \citep{b07}.
This formula assumes a circular orbit and ignores the BH spins. 
We will determine the waveform of the plus mode $h_+(t)$ from the the initial time $(t=0)$ 
when the frequency is $f=10$ Hz to the final time at the inner most stable
 circular orbit corresponding to $f=4400 {\rm Hz} (M_{\rm tot}/{\rm M}_\odot)^{-1}$,
 where $M_{\rm tot}$ is the total mass of the binary.
GWs are lensed by the point mass and the SIS lenses.
The procedure to evaluate the lensed inspiral waveforms is as follows:
(i) prepare the unlensed waveforms $h_+(t)$ in the time domain using the Taylor T4
 formula, 
(ii) perform a Fourier transform\footnote{We used a publicly available code, FFTW3
 (http://www.fftw.org/), to perform the Fourier transform.} to obtain the waveforms in
 the frequency domain $\tilde{h}_+(f)$, and then multiply them by the amplification
 factor $F(w,y)$, and
(iii) finally, repeat the Fourier transform to obtain the lensed waveforms $h^{\rm L}_+(t)$
 in the time domain.
We set the phase of the amplification factor (which can be chosen freely) to be zero
 at the initial frequency $f=10$ Hz, i.e., $T_{\rm GW}(f=10{\rm Hz})=0$.
Therefore, the phases of both the lensed and the unlensed waveforms coincide at $f=10$ Hz.

Figure \ref{lensed_waveforms} shows the lensed and unlensed inspiral waveforms 
 for the $3{\rm M}_\odot+3{\rm M}_\odot$ binary (left panels) and the
 $30{\rm M}_\odot+30{\rm M}_\odot$ binary (right panels) lensed by the point mass
 (upper panels) and the SIS (lower panels).
The red, blue, and green curves are the lensed waveforms with the lens masses
 $1 {\rm M}_\odot,10 {\rm M}_\odot$ and $100 {\rm M}_\odot$ with $y=1$.
The black curves are the unlensed waveforms.
The lower-left subpanels are the results for the initial $0.3$ s.
There are several distinct lensing features in the chirp waveforms, as listed below.

(1) The lensing magnification of the amplitude is increasingly significant as the
 frequency sweeps from low to high.
For example, the lensing magnification is not clearly seen in the small subpanels, but
 becomes clear just before merging.
The magnification is also more significant for the more massive lens because it is
 closer to the geometrical-optics regime.

(2) The phases of the lensed waveforms move slightly to the right as the frequency sweeps
 from low to high due to the GW time delay.
The phase shift is more prominent at higher frequencies, and therefore it can be seen more
 clearly in the left panels ($f=220-700$ Hz) than in the right panels ($f=22-70$ Hz).

(3) Interference patterns appear when two images form in the geometrical-optics
 regime. For example, we can see such modulations in the green curves in
 the left panels: the amplitudes are enhanced at $\sim 300$ Hz, but shrink at
 $\sim 400$ Hz.
 When the path difference between the two images, $c|t_{{\rm d},+}-t_{{\rm d},-}|$, is 
 $n \lambda$ or $(n+1/2) \lambda$ (where $n$ is an integer), the lensed amplitude is
 magnified or reduced due to the interference.

All the lensed waveforms shown in the figure are in the wave-optics regime
 at the initial frequency $f=10$ Hz.
For the green curves in the left panels,
the lensing enters the geometrical-optics regime and therefore the two images form.
From the lensed chirp signals, we can extract information concerning the BH binary
 as well as the lens (its mass and the source position) \citep{tn03}.

\section{Discussion}


We have discussed lensing by a single object.
However, in general, there are multiple lens objects along the line-of-sight to a source.
If the GWs are scattered several times, then the arrival time difference will be
 largely proportional to the number of scatterings.
In this case, the multiple scattering formula in wave optics should be used \citep{y03}.
The multiple scattering calculation is also important to estimate the lensing probability.
However, these calculations are very complicated, and we will discuss them elsewhere.

Let us comment on the LIGO/Fermi GBM transit.
Suppose that the two events are associated and that both signals were
 emitted simultaneously, the measured time lag of $0.4$ s is too large to be explained
 by the lensing of a single object.
The observed frequency range for GW150914 is $f=30-300$ Hz, which corresponds to a
 maximum lensing time lag of $0.3-3$ msec from Eq.(\ref{delta_td_max}).
 { In that case, a lens with $\sim 80 {\rm M}_\odot (f/100 {\rm Hz})^{-1}$ is necessary
   in the line-of-sight to the source but such a probability seems to be quite low.
Therefore, lensing can not explain the measured time lag.}
 

Next, let us comment on the angular position difference between the GW/EM images.
The EM image positions move from their original position due to lensing deflection;
 however, the GW image position does not move in the wave-optics regime.
{ The angular position difference is on the order of the angular Einstein radius $\theta_*$,
which is roughly $10^{-3}$ arcsec for lensing by a galactic star (see Section 2.2) and
 $1$ arcsec for lensing by a galaxy (see Section 3).
However, this deflection is too small to be resolved by current and planned GW detectors.}
Therefore, the image position differences will not be measurable in the near future.

We have shown that GWs can arrive earlier for point mass and SIS lenses.
The GWs always arrive earlier for an arbitrary (positive) lens mass distribution
 because the mass density can be written as a superposition of point masses in
 a weak gravitational field.
Conversely, GWs can arrive {\it later} if they pass through an underdense
 region, such as a void.
The ensemble average of the arrival time differences should be zero
 in a weak gravitational field
 because the mean of the density fluctuations (or the gravitational potential) is
 exactly zero in standard cosmology.



\section{Conclusions}

We have shown that the GW signal arrives earlier than the EM signal if
 they have passed by a lens with mass $\lesssim 10^5 {\rm M}_\odot (f/{\rm Hz})^{-1}$.
This means that the GW propagation is ``superluminal''. 
In general, GWs and EMWs take different times to travel from the source to us when
propagating through an inhomogeneous mass distribution.
{ Therefore, a propagation velocity difference between GWs/EMWs in a vacuum does not
  always mean a breakdown of general relativity.}
{ If a distant SMBHB is lensed by a foreground galaxy, the time difference can reach
  $1-10$ days.
  If future PTAs in the SKA era and optical telescopes can resolve the orbital motions of
  the SMBHBs in both GW/EM signals, they may detect the time lags.
  We also comment that the measured time lag of $0.4$ s in the LIGO/Fermi GBM transit
  (if they are really associated
  and the signals were simultaneously emitted) is too large to be explained by
  the lensing effect.}
The arrival time difference can be used as a ``signal'' to extract the lens information
 if we model the lensed waveform correctly, otherwise it will be a ``noise''.
The arrival time difference could be used as a probe of the cosmic
 abundance of objects with mass $\lesssim 10^5 {\rm M}_\odot (f/{\rm Hz})^{-1}$.
Finally, we can measure the arrival time difference for a {\it single} image
 between multiple frequencies (i.e., multiple images are not necessary).
{ Therefore, the time difference can be measurable even if the waves pass through the
 lens outside the Einstein radius (but not far away).}

Several GW detectors are now in operation, or planned, over a wide frequency range
 from $10^{-8}$ to $10^3$ Hz.
The advanced LIGO-VIRGO network will detect $0.4-400$ NS--NS and $0.2-300$ NS--BH
 mergers per year \citep{a10}.
Some of these mergers will be accompanied by short GRBs.
The Japanese ground-based interferometer KAGRA (the KAmioka GRAvitational wave
 detector) will join this network soon \citep{aso13}.
The space interferometers, eLISA and
 DECIGO (the DECi-hertz Interferometer Gravitational wave Observatory), are scheduled
 to be launched in the 2020s and 2030s, respectively \citep{a12,s01}. 
Pulsar timing arrays are currently in operation and have set an upper bound on the GW
 background at $10^{-8}$ Hz \citep[e.g.,][]{bab16,v16}. 
Inspirals and mergers of supermassive BH (SMBH) binaries are major targets for 
 space-based interferometers and pulsar timing arrays.
These SMBH binaries may be seen by X-ray observations \citep[e.g.,][]{k08,s12,t12}.
In brief, GW astronomy has just begun and a multitude of sources will soon be
 simultaneously detected with both GW and EM detectors.


\acknowledgements

I would like to thank the members of the astrophysics group at Hirosaki University for
 their useful comments and discussions.
I also would like to thank Kunihito Ioka, Teruaki Suyama, Takeshi Chiba and Kazuhiro
 Yamamoto for carefully reading the manuscript and giving me useful comments.
This work was supported in part by MEXT KAKENHI Grant Number 15H05893 and by the JSPS
 Grant-in-Aid for Scientific Research (B) (No. 25287062).



\begin{thebibliography}{}

\bibitem[{Abadie et al.}(2010)]{a10}
 {Abadie}, J., et al. 2010, Class. Quantum Grav., 27, 173001
  
\bibitem[{Abbott et al.}(2016)]{a16}
 {Abbott}, B.P., et al. 2016, \prl, 116, 061102

\bibitem[{Abbott et al.}(2016b)]{a16b}
 {Abbott}, B.P., et al. 2016b, arXiv:1602.03841
 
\bibitem[{Abbott et al.}(2016c)]{a16c}
 {Abbott}, B.P., et al. 2016c, submitted to ApJL, arXiv:1602.08492

\bibitem[{Abbott et al.}(2016d)]{a16d}
 {Abbott}, B.P., et al. 2016d, submitted to ApJS, arXiv:1604.07864

\bibitem[{Amaro-Seoane et al.}(2012)]{a12}
 {Amaro-Seoane}, P., et al. 2012, Class. Quantum Grav., 29, 124016
 
\bibitem[{Andersson et al.}(2013)]{a13}
 {Andersson}, N., et al. 2013, Class. Quantum Grav., 30, 193002

\bibitem[{Aso et al.}(2013)]{aso13}
 {Aso}, Y., et al. 2013, \prd, 88, 043007

\bibitem[{Babak et al.}(2016)]{bab16}
 Babak, S., Petiteau, A., Sesana, A., et al. 2016, \mnras, 455, 1665
 
\bibitem[{Bagoly et al.}(2016)]{b16}
 {Bagoly}, Z., et al. 2016, submitted to A\&A, arXiv:1603.06611

\bibitem[{Baraldo et al.}(1999)]{b99}
 {Baraldo}, C., {Hosoya}, A. \& {Nakamura}, T.T. 1999, \prd, 59, 083001

\bibitem[{Bernardi et al.}(2010)]{b10}
 {Bernardi}, M., {Shankar}, F., {Hyde}, J.B., et al. 2010, \mnras, 404, 2087  
 

\bibitem[{Bliokh} \& {Minakov}(1975)]{bm75}
 {Bliokh}, P.V. \& {Minakov}, A.A. 1975, Astrophys. Space Sci., 34, L7

\bibitem[{Bontz} \& {Haugan}(1981)]{bh81}
 {Bontz}, R.J. \& {Haugan}, M.P. 1981, Astrophys. Space Sci., 78, 199

\bibitem[{Boyle et al.}(2007)]{b07}
 {Boyle}, M., et al. 2007, PRD, 76, 124038

\bibitem[{Burke-Spolaor}(2015)]{b15}
 Burke-Spolaor, S. 2015, arXiv:1511.07869, submitted to PASP
 

\bibitem[{Cao et al.}(2014)]{c14}
 {Cao}, Z., {Li}, L.-F. \& {Wang}, Y. 2014, \prd, 90, 062003
 
\bibitem[{Connaughton et al.}(2016)]{c16}
 {Connaughton}, V., et al. 2016, arXiv:1602.03920

\bibitem[{{Cooray} \& {Seto}}(2004)]{cs04}
 {Cooray}, A., \& {Seto}, N. 2004, \prd, 69, 103502

\bibitem[{{Cutler} \& {Flanagan}}(1994)]{cf94}
 {Cutler}, C., \& {Flanagan}, $\acute{\rm E}$.E. 1994, \prd, 49, 2658 
 
\bibitem[{{Cutler} et al.}(2003)]{chl03}
 {Cutler}, C., {Hiscock}, W.A. \& {Larson}, S.L. 2003, \prd, 67, 024015


\bibitem[{Hobbs et al.}(2010)]{h10}
 Hobbs, G., Archibald, A., Arzoumanian, Z., et al. 2010, Class. Quantum Grav., 27, 084013

\bibitem[{Janssen et al.} (2015)]{j15}
 Janssen, G.H., Hobbs, G., McLaughlin, M., et al. 2015,
 Proceedings of Advancing Astrophysics with the Square Kilometre Array (AASKA14),
 (Giardini Naxos, Italy), 37

\bibitem[{Kahya} \& {Desai} (2016)]{kd16}
 {Kahya}, E.O. \& {Desai}, S. 2016, Phys. Lett. B, 756, 265
 
\bibitem[{Kocsis et al.} (2008)]{k08}
 {Kocsis}, B., {Haiman}, Z. \& {Menou}, K. 2008, \apj, 684, 870

 
\bibitem[{Larson} \& {Hiscock} (2000)]{lh00}
 {Larson}, S.L. \& {Hiscock}, W.A. 2000, \prd, 61, 104008

\bibitem[{Li et al.}(2016)]{l16}
 {Li}, X., et al. 2016, submitted to \apj, arXiv:1601.00180
 
\bibitem[{Macquart}(2004)]{m04}
 {Macquart}, J.P. 2004, A\&A, 422, 761

\bibitem[{Moylan et al.}(2008)]{m08}
 {Moylan}, A., et al. 2008, Proceedings of the MG11 Meeting on General Relativity,
  arXiv:0710.3140

\bibitem[{Nakamura}(1998)]{n98}
  {Nakamura}, T.T. 1998, \prl, 80, 1138
  
\bibitem[{Nakamura} \& {Deguchi}(1999)]{nd99}
  {Nakamura}, T.T., \& {Deguchi}, S. 1999, Progress of Theoretical Physics Supplement, 
  133, 137

  
\bibitem[{Nishizawa} \& {Nakamura}(2014)]{nn14}
  {Nishizawa}, A., \& {Nakamura}, T. 2014, \prd, 90, 044048

\bibitem[{Nishizawa}(2016)]{n16}
  {Nishizawa}, A. 2016, arXiv:1601.01072

\bibitem[{Oguri} \& {Marshall}(2010)]{om10}
 Oguri, M. \& Marshall, P.J. 2010, \mnras, 405, 2579
  
\bibitem[{Oguri et al.}(2012)]{o12}
 Oguri, M., Inada, N., Strauss, M.A., et al., 2012, \aj, 143, 120
  
\bibitem[{Ohanian}(1974)]{o74} {Ohanian}, H.C. 1974, Int. J. Theor. Phys., 9, 425 
  
\bibitem[{Peters}(1974)]{p74} {Peters}, P.C. 1974, \prd, 9, 2207

  
\bibitem[{Rosswog}(2015)]{r15} {Rosswog}, S. 2015, Int. J. Mod. Phys. D, 24, 1530012

\bibitem[{Savchenko et al.}(2016)]{s16} 
 Savchenko, V., Ferrigno, C., Mereghetti, et al. 2016, \apjl, 820, L36 
  
\bibitem[{Schneider et al.}(1992)]{sef92}
  {Schneider}, P., {Ehlers}, J., \& {Falco}, E.E. 1992,
  Gravitational Lenses, Springer, New York

\bibitem[{Sereno et al.}(2010)]{s10}
  Sereno M., Sesana, A., Bleuler, A. 2010, \prl, 105, 251101
  
\bibitem[{Sesana et al.}(2009)]{svv09}
 Sesana, A., Vecchio, A., \& Volonteri, M. 2009, \mnras, 394, 2255

\bibitem[{Sesana et al.}(2012)]{s12}
  {Sesana}, A., {Roedig}, C., {Reynolds}, M.T. \& {Dotti}, M. 2012, \mnras, 420, 860

\bibitem[{Seto et al.}(2001)]{s01}
  {Seto}, N., {Kawamura}, S. \& {Nakamura}, T. 2001, \prl, 87, 221103
  
\bibitem[{Suyama et al.}(2005)]{s05}
  {Suyama}, T., {Takahashi}, R. \& {Michikoshi}, S. 2005, \prd, 72, 043001
  
\bibitem[{Suyama et al.}(2006)]{s06}
  {Suyama}, T., {Tanaka}, T. \& {Takahashi}, R. 2006, \prd, 73, 024026
  
\bibitem[{{Takahashi} \& {Nakamura}}(2003)]{tn03}
  {Takahashi}, R., \& {Nakamura}, T. 2003, \apj, 595, 1039

\bibitem[{Takahashi}(2004)]{t04}
  {Takahashi}, R. 2004, A\&A, 423, 787
  
\bibitem[{Takahashi et al.}(2005)]{tsm05} 
  {Takahashi}, R., {Suyama}, T. \& {Michikoshi}, S. 2005, A\&A, 438, L5
    
\bibitem[{Takahashi}(2006)]{t06}
  {Takahashi}, R. 2006, \apj, 644, 80

\bibitem[{{Tanaka} et al.}(2012)]{t12}
  {Tanaka}, T., {Menou}, K. \& {Haiman}, Z., 2012, \mnras, 420, 705
  
\bibitem[{Verbiest et al.}(2016)]{v16}
  {Verbiest}, J.P.W., et al. 2016, MNRAS, 458, 1267

\bibitem[{Wu et al.}(2016)]{w16}
  {Wu}, X.-F., et al. 2016, arXiv:1602.01566

\bibitem[{Yamamoto}(2003)]{y03}
  {Yamamoto}, K. 2003, astro-ph/0309696
  
\bibitem[{Yoo et al.}(2007)]{y07}
  {Yoo}, C.-M., {Nakao}, K., {Kozaki}, H. \& {Takahashi}, R. 2007, \apj, 655, 691

\bibitem[{Yoo et al.}(2013)]{y13}
  {Yoo}, C.-M., {Harada}, T. \& {Tsukamoto}, N. 2013, 87, 084045

\bibitem[{Yunes et al.}(2016)]{y16}
  {Yunes}, N., {Yagi}, K. \& {Pretorius}, F. 2016, submitted to \prd, arXiv:1603.08955

\bibitem[{{Wang} \& {Smith}}(2011)]{ws11}
  {Wang}, J. \& {Smith}, M.C. 2011, \mnras, 410, 1135
  


\end{thebibliography}

                                                                        
\end{document}